\documentclass[epj]{svjour}
\usepackage{graphicx}
\begin{document}
\title{Links Between Heavy Ion and Astrophysics}
\author{C. J. Horowitz\inst{1} % etc
}                     % Do not remove
\offprints{horowit@indiana.edu}          % Insert a name or remove this line
\institute{Nuclear Theory Center and Department of Physics, Indiana University, Bloomington, IN 47405, USA }
\date{\today }
\abstract{
Heavy ion experiments provide important data to test astrophysical models.    The high density equation of state can be probed in HI collisions and applied to the hot protoneutron star formed in core collapse supernovae.  The Parity Radius Experiment (PREX) aims to accurately measure the neutron radius of $^{208}$Pb with parity violating electron scattering.  This determines the pressure of neutron rich matter and the density dependence of the symmetry energy.  Competition between nuclear attraction and coulomb repulsion can form exotic shapes called nuclear pasta in neutron star crusts and supernovae.  This competition can be probed with multifragmentation HI reactions.  We use large scale semiclassical simulations to study nonuniform neutron rich matter in supernovae.  We find that the coulomb interactions in astrophysical systems suppress density fluctuations.  As a result, there is no first order liquid vapor phase transition.  Finally, the virial expansion for low density matter shows that the nuclear vapor phase is complex with significant concentrations of alpha particles and other light nuclei in addition to free nucleons.        
 % end of PACS codes
} %end of abstract
\maketitle
\section{Introduction}
\label{intro}
Most of the visible mass and energy of the universe is in atomic nuclei.  This suggests some common goals for heavy ion (HI) research.  We can study nuclear matter under extreme conditions of density (both high and low), temperature, size, and isospin.  The insight gained from this study can then be applied to: (1) the fundamental behavior of many particle quantum systems such as cold atoms in laboratory traps, (2) Quantum Chromodynamics at high densities, and (3) compact objects in Astrophysics such as neutron stars, supernovae, gamma ray bursts, accretion disks, and the origin of the chemical elements.  

In this article we discuss links between HI and Astrophysics.  We need to extrapolate HI data to Astrophysical conditions.  First, one must extrapolate to longer times.  Core collapse Supernovae (SN) are giant stellar explosions that produce neutron stars and chemical elements and accelerate cosmic rays.  In SN the core of a massive star collapses in milliseconds.  This is a remarkably short time scale for a planet sized object that is more massive then the sun.  However a msec is $10^{20}$ fm/c! and very long compared to the time scale of a few hundred fm/c for a HI collision.  Therefore, SN involve matter that has had plenty of time to reach thermodynamic equilibrium, while this is not always the case in HI collisions.

Second, one must extrapolate to larger systems.  A neutron star is a giant atom with a mass number of $10^{57}$ and an atomic number of $10^{56}$.  It is about 10 km in radius, or 18 orders of magnitude larger then a conventional atomic nucleus.  For this nearly infinite system, coulomb interactions play a crucial role and require charge neutrality between positively charged nuclear matter and a background electron gas.  Thus, one must consider the differences in coulomb interactions of finite HI collisions compared to those of an infinite system.

Third, one must extrapolate to larger isospin.  Astrophysical systems are often more neutron rich then the heavy ions that are available in the laboratory.  This extrapolation depends on the symmetry energy.  The symmetry energy $S(\rho)$ describes how the energy of nuclear matter rises when one moves away from equal numbers of neutrons and protons.  The density dependence of $S(\rho)$ is very important for many astrophysical systems, and can be determined from HI experiments \cite{hisym}.  Furthermore, future experiments with more neutron rich radioactive beams may provide additional information.

There are errors associated with these extrapolations.  Nevertheless, laboratory HI experiments provide real data that can be used to place important constraints on many Astrophysical models.  Without the HI data, one may be forced to use untested theoretical assumptions that have large errors.          

In this article, we discuss links between HI and Astrophysics. Section II discusses the high density equation of state (EOS) and its implications for neutron star structure and supernovae.  Next, we consider the EOS at sub-nuclear densities.  Section III discusses the Parity Radius Experiment (PREX) to measure the neutron skin thickness in $^{208}$Pb.  This determines the density dependence of the symmetry energy and the neutron matter EOS at low densities.  Section IV presents molecular dynamics simulations of the non-uniform neutron rich matter in the inner crusts of neutron stars.  These nuclear pasta phases may be closely related to multi-fragmentation in HI collisions.  Finally, Section V discusses the nuclear matter liquid-vapor phase transition in supernovae.

\section{The High Density Equation of State}
\label{highdenEOS}

The equation of state (EOS) describes the pressure $P$ of nuclear matter as a function of density $\rho$, temperature $T$, and proton fraction $Y_p$.  Heavy ion experiments can probe the EOS at high $T$ and $\rho$ and for proton fractions near $Y_p\approx 1/2$.  For example, flow observables can be used to constrain the EOS with the help of semiclassical simulations \cite{scienceEOS}.  In addition, yields of other particles such as Kaons can provide additional probes of the EOS \cite{kaonsEOS}.

Unfortunately, it does not appear possible to directly produce cold dense matter in the laboratory.  The energy needed to produce high compression always seems to produce high temperatures because there is no way to get the entropy out.  Therefore Ref. \cite{scienceEOS} assumed the temperature dependence of the EOS was that predicted by some simple mean field models.         

Neutron stars (NS), on the other hand, provide unique probes of the EOS of cold dense matter.  Although they are formed hot in SN explosions, they have plenty of time to cool via neutrino emission.  Thus NS can probe new forms of cold dense matter such as color superconductors that may not be accessible in the laboratory.

It is an exciting time to study neutron stars \cite{scienceNS}.  Powerful X-ray telescopes such as Chandra and XMM-Newton and other instruments are slowly turning NS from theoretical curiosities to detailed, well observed, worlds.  Some NS in binary systems have well measured masses near 1.4 $M_\odot$.  However there are now indications of more massive stars \cite{scienceNS,heavyNS}.  The structure of a neutron star depends only on the EOS of cold neutron rich matter.  The stiffer the EOS (higher pressure for given density), the larger the radius $R(M)$ of a NS, of given mass $M$.  A typical neutron matter EOS may give $R(M)\approx 11-12$ km for $M=1.4M_\odot$, while a stiff EOS could give $R(M)\approx 13-14$ km.  

There is great interest in possible exotic phases for high density matter.  The central density of a NS can be several times nuclear density.  An exotic phase such as strange matter or a color superconductor could lead to a soft high density EOS.  [If the exotic phase has a higher pressure then conventional matter, it may not be thermodynamically favored.]  This could lead to a NS radius of 10 km or less. 

Astronomers are working hard to measure the radii of NS, see for example \cite{radiiNS}.  One approach follows from thermodynamics and the properties of a blackbody radiator.  The luminosity $L$ (total energy radiated per unit time) of an isolated star is related to the surface temperature $T$ and apparent radius $R$ as follows,
\begin{equation} 
L=4\pi R^2 \sigma T^4
\label{R}
\end{equation}
where $\sigma$ is the Stephen Boltzmann constant.  The surface temperature can be deduced from X-ray spectra, while $L$ follows from the apparent magnitude of the star and an accurate measurement of its distance.  Unfortunately, there are a number of complications with this simple formula.  Neutron stars are not perfect black bodies, so corrections from realistic stellar atmosphere models may need to be included.  Interstellar absorption can influence estimates of both $L$ and $T$.  The temperature may not be uniform over the stars surface.  For example $T$ can be larger at the magnetic poles compared to the equator because the thermal conductivity is larger along the magnetic field direction.  The distance to the star may depend on a very delicate measurement of parallax.  Finally, gravity is so strong that the curvature of space is important.  Some light emitted from the far side of the star can be detected and contributes to $L$ because of this curvature.  This increases the apparent radius by of order 30\%.  Nevertheless, astronomers hope to have a number of increasingly accurate measurements of NS radii.  Comparing results form several different NS measurements may provide a good check of these corrections. 

In addition to cold NS, one is also interested in the structure of very young neutron stars as they are being formed in Supernova explosions.  These hot, lepton rich, protoneutron stars can have maximum temperatures as high as 50 MeV.  The EOS of protoneutron stars may be directly related to the EOS deduced from energetic HI collisions because the temperature, density, and proton fraction can be similar.  Furthermore, this protoneutron star EOS is important for SN simulations \cite{simSN}.

\section{The Parity Radius Experiment and the Low Density EOS}
\label{PREX}

We now discuss the EOS at subnuclear densities.  This has many implications for the structure of NS crusts.  One can obtain information on the low density EOS from both HI collisions and from precision measurements on stable nuclei.  The parity radius experiment (PREX) aims to measure the neutron radius of $^{208}$Pb, accurately and model independently, via parity violating electron scattering.  As we discuss below, the neutron radius in Pb determines the density dependence of the symmetry energy and the EOS of low density neutron matter.  This information, from a precision experiment on a stable nucleus, nicely complements the information from HI or radioactive beam experiments.

Parity violation probes neutrons because the weak charge of a neutron is much larger then the weak charge of a proton \cite{donnelly}.  In the standard model the proton weak charge is proportional to the small factor 1-4sin$^2\theta_W$ where $\theta_W$ is the weak mixing angle.  One can isolate weak contributions by measuring the parity violating asymmetry $A$ for elastic electron nucleus scattering.  This is the cross section difference for the scattering of positive $d\sigma/d\Omega_+$ and negative $d\sigma/d\Omega_-$ helicity electrons,
\begin{equation}
A={d\sigma/d\Omega_+ - d\sigma/d\Omega_-\over d\sigma/d\Omega_+ + d\sigma/d\Omega_-}.
\label{asym}
\end{equation}
In Born approximation $A$ is \cite{donnelly}
\begin{equation}
A=\Bigl({G_F Q^2\over 4\pi \alpha 2^{1/2}}\Bigr) {F_W(Q) \over F_{ch}(Q)}
\label{bornasym}
\end{equation}
where $G_F$ is the Fermi constant, $\alpha$ the fine structure constant and $Q$ the momentum transfer.  The charge form factor $F_{ch}(Q)$ is the Fourier transform of the charge density, that is known from electron scattering.  The weak form factor $F_W(Q)$ is the Fourier transform of the weak charge density. This is dominated by the neutron density and thus the neutron density can be deduced from measurements of $A$.  Note, coulomb distortions make $\approx 30$ \% corrections to $A$ for scattering from a heavy nucleus \cite{coulCJH}.  However these can be accurately calculated.      

The Jefferson laboratory PREX \cite{prexweb} aims to measure elastic scattering of 850 MeV electrons from $^{208}$Pb at six degrees in the laboratory.  The goal is to measure $A\approx 0.6$ ppm with an accuracy of 3\%.  This allows the neutron rms radius of $^{208}$Pb to be deduced to 1\%.  A full discussion of the experiment and many possible corrections is contained in the long paper \cite{bigPREX}.

We now discuss the implications of the radius measurement.  Heavy nuclei are expected to have a neutron rich skin.  The thickness of this skin depends on the pressure of neutron rich matter.  The larger the pressure, the larger the neutron radius as neutrons are forced out against surface tension.  Alex Brown showed that there is a strong correlation between the neutron radius in Pb and the EOS of pure neutron matter, as predicted by many different mean field interactions \cite{brown}.  Therefore, the neutron radius in Pb determines $P$ for neutron matter at $\rho\approx 0.1$ fm$^{-3}$.  [This is about 2/3$\rho_0$ and represents some average of the surface and interior density of Pb.] The pressure depends on the derivative of the energy with respect to density.  The energy of pure neutron matter $E_{neutron}$ is the energy of symmetric nuclear matter $E_{nuclear}$ plus the symmetry energy $S(\rho)$,
\begin{equation}
E_{neutron}\approx E_{nuclear} + S(\rho).
\end{equation}
The pressure depends on $dE_{nuclear}/d\rho$ (which is small and largely known near nuclear density $\rho_0$) and $dS(\rho)/d\rho$.    Therefore, {\it the neutron radius in Pb determines the density dependence of the symmetry energy $dS(\rho)/d\rho$ for densities near $\rho_0$.} 

Neutron stars are expected to have a solid neutron rich crust over a liquid interior, while heavy nuclei have a neutron rich skin.  Both the skin of a nucleus, and the NS crust are made of neutron rich matter at similar densities.  The common unknown is the EOS of low density neutron matter.  As a result, we find a strong correlation between the neutron radius of $^{208}$Pb and the transition density of NS crusts \cite{skin_crust}.  The thicker the skin in Pb, the faster the energy of neutron matter rises with density, and the more quickly the uniform liquid phase is favored.  Therefore, a thick neutron skin in Pb implies a low transition density (maximum density) for the NS crust.  

The composition of a neutron star depends on the symmetry energy.  In beta equilibrium the neutron chemical potential $\mu_n$ is equal to that for protons $\mu_p$ plus electrons $\mu_e$, $\mu_n=\mu_p+\mu_e$.  Neutron stars are about 90\% neutrons and 10\% protons plus electrons.  However, a large symmetry energy will favor more equal numbers of neutrons and protons and increase the proton fraction.  Thus, the composition of matter in the center of a neutron star depends on the symmetry energy at high density.

Neutron stars cool by neutrino emission from the interior.  If the proton fraction is large, above about 0.13, then neutrons near the Fermi surface can beta decay to protons and electrons near their Fermi surfaces and conserve both momentum and energy.  This leads to the direct URCA process $n\rightarrow p + e + \bar\nu_e$ followed by $e+p\rightarrow n + \nu_e$ that will efficiently cool a NS by rapidly radiating $\nu\bar\nu$ pairs.  The neutron radius of Pb constrains the density dependence of the symmetry energy near $\rho_0$.  This is the crucial piece of information for extrapolating to find the symmetry energy at large densities.  We find that if the neutron minus proton rms radii in $^{208}$Pb is larger then 0.25 fm, all of the mean field EOS models considered allow direct URCA for a 1.4$M_\odot$ NS \cite{urca}.  Alternatively, if this skin thickness is less then 0.2 fm, none of the mean field models allow direct URCA.

Note, the direct URCA process takes place in the high density interior of a NS at a few or more $\rho_0$.  Therefore, the above relation with the skin thickness in Pb involves an extrapolation to higher density.  Alternatively, energetic HI collisions can directly produce high densities.  Therefore it would be extremely useful if one could infer the high density symmetry energy from HI observables.  Although potentially difficult and model dependent, {\it measuring the symmetry energy at high density is perhaps the single most important HI experiment for the structure of NS}.     

We close this section with a short discussion of other ways to determine the density dependence of the symmetry energy.  If one assumes the symmetry energy depends on a power of the density,
\begin{equation}
S(\rho)\approx S_0 \rho^\gamma,
\end{equation}
then the power $\gamma$ can be approximately related to the skin thickness in $^{208}$Pb as follows,
\begin{equation}
<r_n^2>^{1/2}-<r_p^2>^{1/2} \approx 0.22 \gamma + 0.06\ {\rm fm}.
\end{equation}
This relation is a simple fit to several mean field calculations, see also \cite{steiner}.  As discussed by Li et al \cite{li} and by Colonna and Tsang \cite{tsang} in the chapter on isospin properties of this book, the power $\gamma$ can be deduced from HI data involving observables such as isoscaling and isospin diffusion.  Finally we mention a recent review article which discusses the symmetry energy in astrophysics \cite{physreports}.

\section{Nuclear Pasta and Multifragmentation}

Nuclei involve an important interplay between Coulomb and nuclear interactions.  Indeed, all baryonic matter is {\it frustrated}.  Nucleons tend to be correlated at short distance, because of short range nuclear attraction, and anti-correlated at long distances because of Coulomb repulsion.  Normally, the nuclear and atomic (or Coulomb) length scales are well separated so nucleons bind into nuclei that are segregated on a crystal lattice.    

However at densities just below $\rho_0$, in the inner crust of neutron stars and in supernovae, coulomb and nuclear scales become comparable.  Under these conditions, the surface energy, from nuclear attraction that favors spherical shapes, and the coulomb energy, that can favor non-spherical shapes, compete.  This results in exotic nuclear pasta phases \cite{nucpasta} that can involve spherical (meat ball), rod (spaghetti), plate (lasagna), or other shapes.            

The coulomb frustration in nuclear pasta is similar to the frustration found in many condensed matter systems.  Frustrated systems can not satisfy all of their elementary interactions \cite{frust}.  Examples range from magnetism \cite{magfrust} to protein folding \cite{fold}.  Because frustration raises the energy of the ground state, these systems are characterized by a very large number of low energy excitations that lead to unusual dynamics.

Nuclear pasta may be important for a number of neutron star observables.  For example, r-modes are collective oscillations of NS that can radiate gravitational waves and may control pulsar spin periods \cite{rmodes}.  The sheer viscosity of the nuclear pasta at the interface between the solid crust and liquid interior of a NS may determine the damping of r-modes.  This viscosity in turn may depend crucially on the exotic shapes of the pasta.  Some other relevant pasta properties include thermal conductivity, sheer modules, and neutrino emissivity.   

Core collapse supernovae radiate of order 10$^{58}$ neutrinos.  The very large gravitational binding energy of the newly formed neutron star (100 to 200 MeV/A) is released, almost entirely, in neutrinos.  No other known particles can transport the energy out of the very dense core during the few second duration of the explosion.  These 10 to 20 MeV neutrinos can scatter coherently from the nuclear pasta because their wave lengths are comparable to the sizes of the pasta shapes.  Thus, neutrino-pasta scattering \cite{pasta} may be important for supernova dynamics. 

Nuclear pasta in astrophysics may be closely related to multifragmentation in laboratory heavy ion collisions.  Heavy ions, at moderate excitation energy, are observed to break apart into several large fragments \cite{multifragmentation}.  This process may occur at the same, slightly subnuclear, densities where nuclear pasta forms.  Furthermore, both pasta formation and multifragmentation are driven by the same nuclear and coulomb energies.  One may be able to tune the interactions used in semiclassical simulations of multifragmentation, in order to reproduce laboratory data.  Then, the same simulations and interactions can be used to describe nuclear pasta.   This allows laboratory data to be used to constrain astrophysical models.

It is important to go beyond mean field models in describing nuclear pasta.  Mean field interactions, fit to conventional nuclei, may not be appropriate for complex non-uniform pasta.  Furthermore, pasta may not be described well by a Maxwell construction, such as in ref. \cite{serot} involving uniform liquid and uniform gas phases.  In addition, the coulomb interaction plays a crucial role in astrophysics.  The system must be electrically neutral.  Therefore, the positive charge density of the pasta is constrained to be equal and opposite to the electron density.  Finally, one should consider a wide variety of possible shapes for the nuclear pasta.  Variational calculations involving a few simple shapes, such as rods or plates, may miss more complicated configurations. 

In ref. \cite{pasta} we consider a simple semiclassical model where neutrons and protons interact via short ranged nuclear and screened coulomb forces.  The electrons form a very degenerate Fermi gas and are not included explicitly.  Instead, the very slight polarization of the electrons lead to a Thomas Fermi screening length $\lambda$ for the Coulomb interactions between protons.  Our model Hamiltonian is
\begin{equation}
H=\sum_i \frac{p_i^2}{2m} + \sum_{i<j} V(i,j)
\end{equation}
where the two body potential is
\begin{equation}
V(i,j)=a{\rm e}^{-r_{ij}^2/\Lambda}+ [b+c\tau_z(i)\tau_z(j)]{\rm e}^{-r_{ij}^2/2\Lambda} + V_c(i,j).
\end{equation}
Here the distance between the particles is $r_{ij}=|{\bf r}_i-{\bf r}_j|$ and the isospin of the $j_{th}$ particle is $\tau_z(j)=1$ for a proton and $\tau_z(j)=-1$ for a neutron.  The model parameters $a$, $b$, $c$, and $\Lambda$ have been fit to reproduce the binding energy and saturation density of nuclear matter along with a reasonable symmetry energy \cite{pasta}.  The screened Coulomb interaction is
\begin{equation}
V_c(i,j)=\frac{e^2}{r_{ij}}{\rm e}^{-r_{ij}/\lambda}\tau_p(i)\tau_p(j)
\end{equation}
where $\tau_p(j)=(1+\tau_z(j))/2$ is the nucleon charge and $\lambda$ is the screening length from the slight polarization of the electrons.  

This model yields large nuclei or pieces of pasta that are heavy and have thermal Compton wavelengths much shorter than their inter-particle spacing.  This motivates our semiclassical approximation.  More elaborate interactions can be employed, such as the QMD calculations of Watanabe et al. \cite{watanabe}.  However our simple interaction reproduces nuclear saturation and includes Coulomb interactions.  We believe these are the most important features that determine the long range structure of the nuclear pasta phases. 

The wavelength of a 10 MeV supernova neutrino is 120 fm.  To determine the pasta structure at this long length scale may require simulations involving many particles.  For example, at $1/3\rho_0$ there are 100,000 nucleons in a cube 120 fm on a side.  We have used special purpose MDGRAPE computer hardware to perform molecular dynamics simulations with 40,000 to 200,000 nucleons \cite{pasta2}.

We are interested in the neutron rich matter during a supernova.  The proton fraction starts near 1/2 and drops to low values as electron capture proceeds and electron neutrinos diffuse out of the core.  Figure 1 shows a sample configuration of 40,000 nucleons at a density of 0.01 fm$^{-3}$, a proton fraction of 0.2, and a temperature of $T=1$ MeV.  An iso-surface of the proton density is shown.  At this density, most of the protons cluster into neutron rich nuclei.   Between these nuclei, there is a low density neutron gas that is not shown in Fig. 1.       

\begin{figure}
\centering
  \includegraphics[width=2.75in]{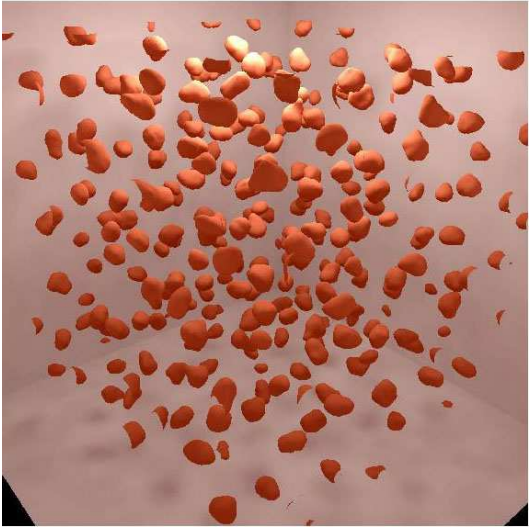}
\caption{Proton density iso-surface for a sample configuration of 40,000 nucleons at $\rho=0.01$ fm$^{-3}$, $T=1$ MeV and proton fraction 0.2.  The simulation volume is about 160 fm on a side.}
\label{fig:1}       % Give a unique label
\end{figure}
   
To characterize the heavy nuclei in Fig. 1 we have used a clustering algorithm.  A nucleon is said to belong to a cluster if it is within a cutoff radius $R_{C}\approx 3$ fm of at least one other nucleon in the cluster.  This divides the 40,000 nucleons into about 12,000 free neutrons, a collection of light nuclei, and about 250 heavy nuclei as shown in Fig. 2.  The heavy nuclei have an average mass near $<A>\approx 100$ and a $Z/A\approx 0.3$.  Note, this $Z/A$ is somewhat greater then the total proton fraction of 0.2 because of isospin distillation.  The rest of the neutrons go into the low density neutron gas.  Our simulation results are qualitatively similar to many statistical models such as those of Botvina \cite{botvina}.  The distribution of clusters reflects a balance between binding energy, favoring large clusters, and entropy, that favors light clusters.  However in detail, the distribution can be sensitive to the nuclear masses predicted by a given model.

\begin{figure}
%\resizebox{0.75\columnwidth}{!}{%
\centering
 \includegraphics[width=2.75in,angle=270]{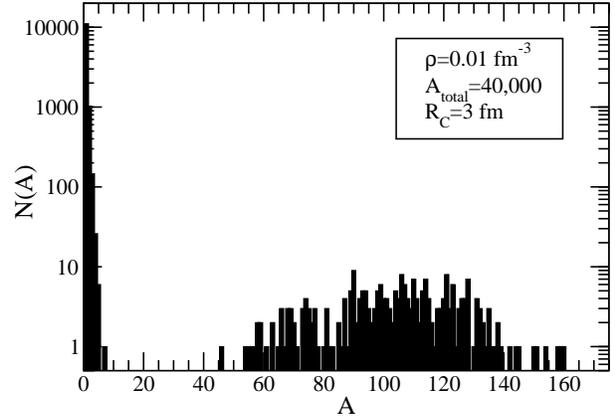}
%}
\caption{Fragment size distribution for the sample configuration of Fig. 1, see text.}
\label{fig:2}       
\end{figure}
   
As the density increases, the background electron gas cancels more of the Coulomb interaction.  This allows the formation of larger clusters.  In Fig. 3 we compare the cluster distribution at $\rho=0.01$ fm$^{-3}$ to that at $\rho=0.025$ fm$^{-3}$ (for the same $T=1$ MeV and proton fraction 0.2).  At $\rho=0.025$ fm$^{-3}$ the average mass is now $<A>\approx 200$ and there is a tail in the distribution to very heavy nuclei.      

\begin{figure}
\centering
\includegraphics[width=2.75in,angle=270]{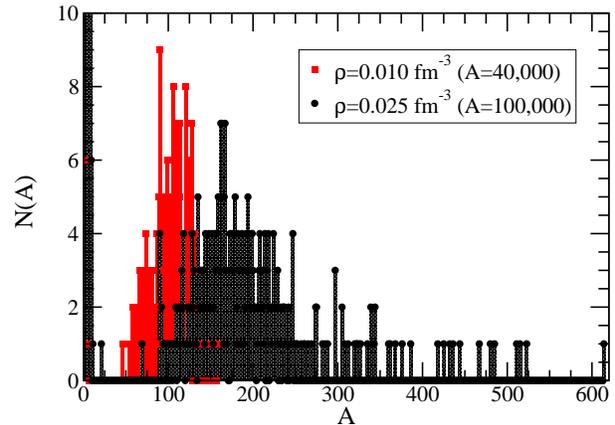}
\caption{Fragment size distributions at $\rho=0.01$ and 0.025 fm$^{-3}$.}
\label{fig:3}       
\end{figure}

Finally, as the density is increased further the nuclei start to strongly interact.  Figure 4 shows an is-surface of the proton density at $\rho=0.05$ fm$^{-3}$ ($\approx 1/3\rho_0$).  Now spherical nuclei are no longer favored.  Instead, long spaghetti like strands are seen that have complex shapes.  The fragment distribution now includes very large clusters whose size scales with the simulation volume.  Thus, heavy nuclei have percolated together to form a complex pasta phase.  Note that increasing the density still further to $\rho=0.075$ fm$^{-3}$ ($1/2\rho_0$) results in a transition to uniform nuclear matter, not shown.
 
\begin{figure}
\centering
  \includegraphics[width=2.75in]{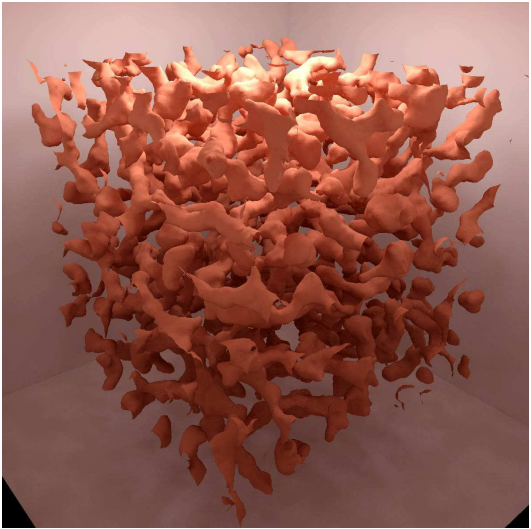}
\caption{Proton density iso-surface for a sample configuration of 100,000 nucleons at $\rho=0.05$ fm$^{-3}$, $T=1$ MeV and proton fraction 0.2.  The simulation volume is about 120 fm on a side.}
\label{fig:4}    
\end{figure}

The clusters seen in Figs 1 and 4 can be characterized by the static structure factor $S_q$ \cite{pasta,pasta2}.  This describes the degree of coherence for neutrino scattering from the nonuniform system.  This is directly analogues to $S_q$ for many complex condensed matter systems that can be deduced from neutron or X-ray scattering.  The static structure factor coherently sums the reflected waves for neutrino scattering from each neutron in the system,
\begin{equation}
S_q=\sum_{i,j} {\rm exp}[i{\bf q}\cdot ({\bf r}_i-{\bf r}_j)],
\end{equation}
where ${\bf q}$ is the momentum transferred from the neutrino to the system.
In Fig. 5 we show $S_q$ for densities of 0.01, 0.025, 0.05, and 0.075 fm$^{-3}$.  This scans the density range from largely isolated nuclei (in Fig. 1) through the complex pasta phases (Fig. 4) to uniform nuclear matter.  A large peak is seen in $S_q$ for $q\approx 0.3$ fm$^{-1}$.  This corresponds to neutrino nucleus elastic scattering at $\rho=0.01$ fm$^{-3}$ or coherent neutrino-pasta scattering at $\rho=0.05$ fm$^{-3}$.  Here the neutrino scatters coherently from all of the neutrons in a cluster.  This peak largely vanishes for the uniform system at $\rho=0.075$ fm$^{-3}$.

\begin{figure}
\centering
  \includegraphics[width=2.75in,angle=270]{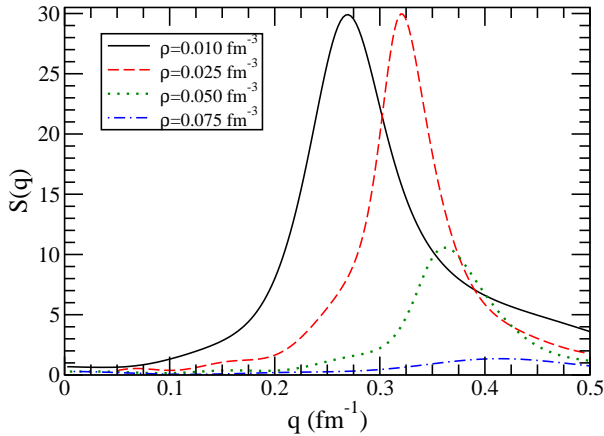}
\caption{The static structure factor $S_q$ for $T=1$ MeV and $Y_p=0.2$ for the indicated densities.}
\label{fig:5}    
\end{figure}

At low $q$, $S_q$ is small in Fig. 5 because of ion screening.  If one places an impurity heavy nucleus or piece of pasta into the system, the other clusters will rearrange because of Coulomb interactions until they act to screen the charge of the impurity.  This leads to a reduction of $S_q$.  In the next section, we will use these results for $S_q$ to discuss the liquid vapor transition.

One can use the time dependence of the molecular dynamics simulations to calculate the dynamical response function $S(q,w)$ that measures how likely it is for a neutrino to transfer momentum $q$ and energy $w$ to the system.  At $\rho=0.05$ fm$^{-3}$, we find a high energy peak in $S(q,w)$ that represents plasma oscillations of the charged pasta and a peak at low $w$ that may correspond to nucleons diffusing between the pasta and the vapor \cite{response}.  
 
\section{Liquid-Vapor Transition}

There is great interest in the transition between a nucleon vapor at low densities and liquid nuclear matter at high density, see for example \cite{phasetransition}.  Often this is described as a first order phase transition.  However, here we would like to discuss two complications to this simple first order picture that arise in the thermodynamic limit.  First, we believe the low density vapor must necessarily be complex and involve heavier nuclei such as alpha particles in addition to free nucleons.  Second, coulomb interactions replace a first order liquid-vapor phase transition with complex mixed phases such as nuclear pasta.

\begin{figure}
\centering
  \includegraphics[width=2.75in,angle=270]{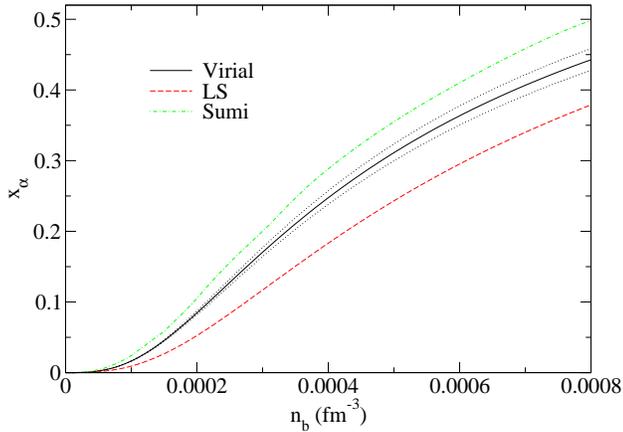}
\caption{The alpha particle mass fraction $X_\alpha$ in symmetric nuclear matter versus density at a temperature of 4 MeV as calculated in the Virial expansion (solid curve).  The error bars are from estimates of the neglected third virial coefficients.  The curves labeld LS and Sumi are from phenomenological models, see \cite{virialnuc}.}
\label{fig:6}    
\end{figure}

The vapor phase, in the limit of very low densities, can be described exactly with the Virial expansion \cite{virialneut,virialnuc}.  Here, the pressure $P$ is expanded in powers of the fugacity $z=$exp$(\mu/T)$ where $\mu$ is the chemical potential.  The second virial coefficient $b_2$, that gives the $z^2$ contribution to the pressure, can be calculated exactly in terms of the two-body elastic scattering phase shifts.  However, nuclear matter is self-bound and tends to form clusters, see Fig. 1.  In ref. \cite{virialnuc} we considered a system of neutrons, protons, and alpha particles.  Because of their large binding energy, alphas tend to be more important then mass 3 nuclei.  Furthermore at very low densities, heavy nuclei are disfavored because of their low entropy.  We calculated the relevant second virial coefficients from $NN$, $N-\alpha$, and $\alpha-\alpha$ elastic scattering phase shifts.  This allows one to make model independent predictions for the alpha particle fraction in the low density vapor, see Fig. \ref{fig:6} \cite{virialnuc}.  Errors in this fraction can be estimated from neglected third Virial coefficients.  

The alpha fraction can be large.  Therefore, even at very low densities say $0.001\rho_0$, the vapor, in the thermodynamic limit, must contain more then just free nucleons.  Note that the virial expansion is exact in the limit of very low density.  It shows that the alpha fraction is nonzero and grows with increasing density, without having to pass through a phase transition.           

It is interesting to compare this complex nuclear vapor to steam in the $H_2O$ system.  This may be the model for a liquid vapor phase transition.  Clusters of multiple $H_2O$ molecules do indeed form, see for example \cite{h2oclusters}.  However their abundance is very low.  In contrast, the large alpha binding energy leads to much larger alpha concentrations.  Therefore nuclear vapor may be much more complex then water vapor.   
 
We now discuss a possible first-order liquid vapor phase transition in astrophysics.  A two phase coexistence region has large density fluctuations as low density vapor is converted to or from a high density liquid.  Scattering from these fluctuations could greatly reduce the neutrino mean free path in a supernova \cite{margueron}.

The static structure factor, in the long wave length limit, $S_{q=0}$ describes fluctuations in the number of neutrons $N$ or density fluctuations,
\begin{equation}
S_{q=0}=\frac{1}{N}(\langle N^2\rangle-\langle N \rangle^2).
\end{equation}
If we assume fluctuations in the neutron density are proportional to fluctuations in the baryon density, this can be written,
\begin{equation}
S_{q=0}\approx(\frac{N}{N+Z})\frac{T}{dP/dn}.
\end{equation}
When two phases coexist, the pressure is constant at the vapor pressure, and the derivative of the pressure with respect to density vanishes $dP/dn=0$.  Therefore, $S_{q=0}$ diverges in a two phase coexistence region of a first order liquid vapor phase transition.

However we find in Fig. \ref{fig:5} that $S_{q=0}$ is small, from ion screening, instead of diverging from density fluctuations.  {\it Therefore, the system does not undergo a first order liquid vapor phase transition.}  The complex structures seen in Fig. \ref{fig:4} can be viewed as a mixed phase with the positively charged nuclear pasta "liquid" in equilibrium with a low density nucleon vapor that occupies the space between the pasta, and is not shown in Fig. \ref{fig:4}.  However the average charge density of the pasta must be equal and opposite to the background electron charge density. Therefore coulomb interactions suppress density fluctuations and eliminate a first order liquid vapor phase transition. 

Note, coulomb interactions for the relatively small system of a heavy ion collision, may be smaller and still allow features of a liquid-vapor phase transition.  However, coulomb interactions, in the nearly infinite astrophysical system, may play a larger role suppressing density fluctuations and modifying the liquid-vapor phase transition.   

\section{Summary}

Heavy ion experiments provide important data to test astrophysical models.  In general, one must extrapolate HI data to longer times, larger sizes, and more neutron rich systems.  The high density equation of state can be probed in HI collisions and applied to the hot protoneutron star formed in core collapse supernovae.  The Parity Radius Experiment (PREX) aims to accurately measure the neutron radius of $^{208}$Pb with parity violating electron scattering.  This determines the pressure of neutron rich matter and the density dependence of the symmetry energy.  Competition between nuclear attraction and coulomb repulsion can form exotic shapes called nuclear pasta in neutron star crusts and supernovae.  This competition can be probed with multifragmentation HI reactions.  A first order liquid vapor phase transition has density fluctuations that could impact neutrino interactions in supernovae.  We use large scale semi-classical simulations to study non-uniform neutron rich matter.  We find that the coulomb interactions in astrophysical systems suppress density fluctuations. As a result, the system does not undergo a first order liquid vapor phase transition.  Finally, the virial expansion for low density matter shows that the nuclear vapor phase is complex with significant concentrations of alpha particles and other light nuclei in addition to free nucleons.        

\section{Acknowledgements}
Collaborators for this work include Achim Schwenk, Jorge Piekarewicz, Angeles Perez-Garcia, and Don Berry.  We thank Marcello Baldo for suggestions.  We acknowledge financial support form the U. S. Department of Energy contract DE-FG02-87ER40365.

%\begin{figure*}
%\vspace*{5cm}       % Give the correct figure height in cm
%\end{figure*}
%\begin{table}
%\begin{tabular}{lll}
%\hline\noalign{\smallskip}
%\end{tabular}
% Or use
%\vspace*{5cm}  % with the correct table height
%\end{table}
%
% BibTeX users please use
% \bibliographystyle{}
% \bibliography{}

\end{document}